\documentclass[a4paper]{jpconf}
\usepackage{graphicx}
\usepackage{iwsmihead}

\let\a=\alpha   \let\g=\gamma  \let\d=\delta \let\e=\varepsilon
     \let\l=\lambda
                
 \let\t=\tau    
   
 \let\D=\Delta   
    \let\Si=\Sigma     
  \let\r=\rho 
\let\io=\infty
\def\ie{{i.e. }}

\def\NN{{\cal N}}

\def\to{\rightarrow}
\def\la{\left\langle}
\def\ra{\right\rangle}

\newcommand{\beq}{\begin{equation}}
\newcommand{\eeq}{\end{equation}}

\newcommand{\wt}{\widetilde}

\begin{document}

\title{Relationship between clustering and algorithmic phase transitions in the
random k-XORSAT model and its NP-complete extensions}

\author{Fabrizio Altarelli$^{1,2}$, R\'emi Monasson$^2$ and
Francesco Zamponi$^{2,3}$}

\address{
$^1$ Dipartimento di Fisica, Universit\`a di Roma ``La Sapienza'', P.le A. Moro
2, 00185 Roma, Italy \\
$^2$ CNRS-Laboratoire de Physique Th\'eorique, Ecole Normale Sup\'erieure, 24
rue Lhomond, 75005 Paris, France \\
$^3$ Service de Physique Th\'eorique, Orme des Merisiers, 
CEA Saclay, 91191 Gif-sur-Yvette Cedex, France
}

\ead{fabrizio.altarelli@roma1.infn.it}

\date{\today}
\begin{abstract}
We study the performances of stochastic heuristic search algorithms on Uniquely
Extendible Constraint Satisfaction Problems with random inputs. We
show that,  for any heuristic
preserving the Poissonian nature of the underlying instance, 
the (heuristic-dependent) largest ratio  $\alpha _a$ of constraints
per variables for which a search algorithm is likely to find solutions
is smaller than the critical ratio $\alpha_d$ above
which solutions are clustered and highly correlated. In
addition we show that the clustering ratio can be reached when the number
$k$ of variables per constraints goes to infinity by the so-called
Generalized Unit Clause heuristic.
\end{abstract}

\section{Introduction}

The application of statistical mechanics ideas and tools to random optimization
problems, initiated in the mid-eighties \cite{Mezard87}, has benefited from a
renewed interest from the discovery of phase transitions in Constraint
Satisfaction Problems (CSP) fifteen years ago. Briefly speaking, one wants to
decide whether a set of randomly drawn constraints over a set of variables
admits (at least) one solution. When the number of variables goes to 
infinity at
fixed ratio $\alpha$ of constraints per variable the answer abruptly changes
from (almost surely) Yes to No when the ratio crosses some critical value
$\alpha_s$. Statistical physics studies have pointed out the existence of
another phase transition in the Yes region \cite{Biroli00, Krzakala07}. The set
of solutions goes from being connected to a collection of disconnected clusters
at some ratio $\alpha_d < \alpha_s$, a translation in optimization terms of the
replica symmetry breaking transition identified by Parisi in mean-field spin
glass theory.

It is expected that this clustering transition may have dynamical consequences.
As replica symmetry breaking signals a loss of ergodicity, sampling algorithms
(e.g. Monte Carlo procedure) run into problems at that transition. A
quantitative study of the slowing down of MC scheme was done in
\cite{Montanari06} for the case of the $k$-XORSAT model where constraints are
simply linear equations (modulo 2) over $k$ Boolean variables (for an
introduction, see \cite{Monasson07} and references therein). Yet, finding a
solution should in principle be easier than sampling, and the exact nature of
the relationship between the performances of resolution algorithms and the
static phase transitions characterizing the solution space is far from being
obvious \cite{Krzakala07a}. The present paper is a modest step in elucidating
this question for the $k$-XORSAT problem, and some related NP-complete
problems sharing the same random structure.

Hereafter we consider simple stochastic search heuristic algorithms working in
polynomial (linear) time for solving $k$-XORSAT instances
\cite{Chao90,Monasson07}. By successively assigning variables
according to some heuristic rules those 
algorithms either produce a solution, or end up with a
contradiction. The probability that a solution is found is a
decreasing function of the ratio $\alpha$, and vanishes above some
heuristic-dependent ratio $\alpha_a$ in the infinite size limit.
We show that $\alpha_a < \alpha_d$ for any assignment heuristic in the
class of rules preserving the Poissonian structure of the instance.
In addition, we determine the most efficient heuristic, that is, the one
maximizing $\alpha_a$ in this class and show that for large $k$, the
two critical ratios match, $\alpha _a(k) \simeq \alpha_d(k) \simeq
\log k/k$.

The plan of the paper is as follows. In section \ref{sec:def} we
define the random $k$-XORSAT decision problem
and its extension, as well as  the search 
algorithms studied. Section \ref{sec:leaf}
presents a method to characterize the phase diagrams of those random decision 
problems, depending on the content (numbers of
constraints over $j$ variables, with $j$ ranging from 1 to $k$)
of their instances. We show that all
important information is encoded in a unique `thermodynamical'
potential for the fraction of frozen variables (backbone). The
analysis of the dynamical evolution of the instance content is
exposed in section \ref{sec:dyna}. These dynamical results are
combined with the static phase diagram in section \ref{sec:cinq} to
show that the success-to-failure critical ratio of search heuristic,
$\alpha_a$, is smaller than the ratio corresponding to the onset of
clustering and large backbones, $\alpha_d$. We then show that 
 the so-called Generalized Unit Clause heuristic rule is optimal (in
 the class of Poissonian heuristics) and its critical ratio $\alpha_a$
is asymptotically equal to $\alpha_d$ in the large $k$ limit. Our
results are discussed in section \ref{sec:conc}.

\section{Definitions}
\label{sec:def}

\subsection{Decision problems}

The decision problems we consider in this paper are $(k,d)$-Uniquely Extendible (UE)
Constraint Satisfaction Problems (CSP)
defined as follows~\cite{Connamacher04}.  One considers $N$ variables
$x_i \in \{0,1,\cdots,d-1\}$.  A UE constraint,
or {\it clause}, is a constraint on $k$ variables such that, if one
fixes a subset of $k-1$ variables, the value of the $k$-th variable is
uniquely determined.  A $(k,d)$-UE-CSP formula is a collection of $M =
\a N$ clauses, each involving $k$ variables (out of the $N$ available
ones). A {\it solution} is an assignment of the $N$ variables such that all the
clauses are satisfied.
$k$-XORSAT corresponds to $d=2$ and is solvable in polynomial time
with standard linear algebra techniques. For $d=3$ the problem is
still in P, while for $d\ge 4$ it has been shown that $(3,d)$-UE-CSP is NP-complete
\cite{Connamacher04}.

A random formula is obtained by choosing, for each clause, the
$k$ variables, and the actual UE constraint, uniformly at random.
It is known that, in the infinite size limit
$N\to\infty$ and at fixed clause-to-variable ratio $\alpha$, 
\cite{Connamacher04,Dubois02,Mezard03,Cocco03}:
\begin{itemize}
\item there is a critical ratio $\alpha _s (k)$ such that a random 
$(k,d)$-UE-CSP is almost surely satisfiable (respectively,
  unsatisfiable) if $\alpha < \alpha_s(k)$ (respectively, $\alpha >
  \alpha _s (k)$).
\item in the satisfiable phase there is
  another phase transition at some ratio $\alpha _d (k)$ such that:
\subitem{-} for $\alpha < \alpha_d(k)$ the space of solutions is
  `connected': with high probability there is a path in the set of
  solutions joining any two solutions such that a step along the path
  requires to change O(1) variables.
\subitem{-} for $\alpha > \alpha _d(k)$ the space of solution is
  disconnected into an exponentially large number of clusters, each
  one enjoying the above connectedness property, and far away from
  each other (going from one solution in one cluster to another
  solution in another cluster requires to change $O(N)$ variables).
In addition, in each cluster, a finite fraction of variables are
  frozen {\em i.e.} take the
  same value in all solutions (backbone).
\end{itemize}

\subsection{Search algorithms}

We will consider simple algorithms acting on the formula in an attempt to find
solutions. Those algorithms were introduced and analyzed by Chao and Franco
\cite{Chao90} (see \cite{Achlioptas01} for a review). Briefly
speaking, starting
from a randomly drawn formula, the algorithm assigns one variable at each time
step  according to the following principles: 
\begin{itemize}
\item If there is (at least) one clause of length one (called
  unit-clause) then satisfy it by adequately assigning its
  variable. This rule is called {\em unit propagation}.
\item If all clauses have length two or more, then choose a variable
  according to some heuristic rules. Two simple rules are:
\subitem{-} Unit Clause (UC): pick up uniformly at random any variable
  and set it to a random uniform value in $\{0,\cdots,d-1\}$;
\subitem{-} Generalized Unit Clause (GUC): pick up uniformly at random
one of the shortest clauses, then a variable it this clause, and
  finally its value.
\end{itemize}
In this analysis, we will discuss a general heuristics in which the variable to
be set is chosen among those that appear in the clauses of length $j$ with some
probability $p_j(C_1,\cdots,C_k)$, depending in general on the number of clauses
of length $j$ present in the formula, that we shall call $C_j$.
Unit propagation implies that if $C_1 \neq 0$, then $p_j = \d_{j,1}$.
We consider also the possibility that the variable is chosen irrespective of
the clause length, then $\sum_{j=1}^k p_j \leq 1$.

Both UC and GUC are special cases of this general class: in UC variables are
chosen at random, irrespectively of the clauses they appear in (if any), so that
$p_j = 0$ unless there are unit clauses; GUC corresponds to $p_j = \d_{j,j^*}$
where $j^*$ is the length of the shortest clause in the system. Notice that
since the variables are selected independently of their number of occurrences,
the latter remains Poissonian under the action of the algorithm (even though the
value of the parameter in the distribution of occurrences may vary). More
involved heuristics do exist but will not be analyzed here.

Under the action of the algorithm clauses get reduced (decrease in length) until
they disappear once satisfied. The algorithm stops either when all clauses have
been satisfied or when two incompatible unit-clauses have been generated e.g.
$x=0$ and $x=1$. In the latter case the algorithm outputs `I do not know whether
there is a solution', while in the former case the output reads
`Satisfiable' and returns a solution to the formula. 
The probability of success, that is, the probability (over the
choices of the algorithms and the formula) of getting the `Satisfiable' output
vanishes above some heuristic-dependent ratio $\alpha_a (< \alpha_s)$ in the
infinite $N$ limit. This success-to-failure transition coincides with
the polynomial-to-exponential transition of backtracking algorithms
\cite{Monasson07,Achlioptas04}.

\section{`Thermodynamical' Characterization of the Space of Solutions}
\label{sec:leaf}

Under the action of the algorithm the length of the clauses changes; therefore
the initial $(k,d)$-UE-CSP formula where all clauses have length $k$ 
evolves into a formula with some distribution of clauses of different lengths.
We wish then to characterize the space of solutions of a generic $d$-UE-CSP
formula made by $N$ variables and by $\{C_j^0\}_{j=2,\cdots,k}$ clauses of
length $j$, assuming that there are no unit clauses. This characterization will
be useful to analyze the performance of search algorithm in the following.

\subsection{Leaf removal procedure and its analysis}

Our starting observation is that, due to the UE property, when a variable has a
unique occurrence in the formula, then the clause it appears in can always be
satisfied. Hence the subformula obtained by removing this clause is equivalent
(in terms of satisfiability) to the original system~\cite{Dubois02}. The
interest of this remark is that it can be iterated, and more and more clauses
eliminated. Monitoring the evolution of the formula under this procedure, called
leaf removal, provides us with useful information on the nature of the solution
space \cite{Mezard03, Cocco03, Weigt02}.

One clause is removed at each time step. After $T$ steps we denote by $C_j(T)$
the number of clauses of length $j$. Those numbers obey the evolution equations
(in expectation), 
\beq \label{eq_gen_disc}
C_j(T+1)-C_j(T) = -\frac{j\;C_j(T)}{\sum_{j'=2}^k j' C_{j'}(T)} 
\eeq
where the denominator is the total number of occurrences of all variables
appearing in the formula. The r.h.s. of (\ref{eq_gen_disc}) is simply (minus)
the probability that the unique-occurrence variable is drawn from a clause of
length $j$.

In addition let us define the number $N_\ell(T)$ of variables appearing in
$\ell$ equations exactly. The evolution equations for those numbers are (in
expectation)
\beq\label{eq_gen_disc2}
N_\ell(T+1)-N_\ell(T) =  \sum_{j=2}^k \frac{j(j-1)\;C_j (T)}{\sum_{j'=2}^k j'
C_{j'}(T)} 
\times \left[ \frac{(\ell+1)\;
N_{\ell+1}(T)-\ell\;N_\ell(T)}{\sum_{\ell'=0}^\infty \ell' N_{\ell'}(T)}
  \right] \ - \d_{\ell,1} + \d_{\ell,0} \ .
\eeq 
The above is easy to interpret. The second term in the square bracket on the
r.h.s. is the average number of removed variables (other than the
single-occurrence variable), that is, the average length of the removed clause
minus one. The first term expresses that, if one of those variables appeared
$\ell+1$ times before its removal, the number of its occurrences has decreased
down to $\ell$ after the removal. Finally, the two $\d$ correspond to the
elimination from the system of the single-occurrence variable.

In the large $N$ limit we may turn those finite difference equations over
extensive quantities $C_j,N_\ell$ into differential equations for their
intensive counterparts $c_j=C_j/N, n_\ell =N_\ell/N$ as functions of the reduced
number of steps, $\tau =T/N$. The outcome is
\begin{eqnarray}\label{LRc}
  \frac{d c_j}{d \t} &=& - \frac{jc_j}{\NN} \ , \ \ \ (j=2,\dots,k) \ , \\
\label{LRn}
  \frac{d n_\ell}{d \t} &=& \sum_{j=2}^k \frac{j (j-1) c_j}{\NN}  \left[
        \frac{(\ell+1)n_{\ell+1}-\ell n_\ell}{\NN} \right] - \delta_{\ell,1} +
\delta_{\ell,0} \ ,
\end{eqnarray}
where $\NN(\t) = \sum_{j=2}^k j c_j(\t) = \sum_{\ell\ge 1} \ell\; n_\ell(\t)$.
The initial conditions are 
\beq\label{Poissonian}
c_j (0) = \frac{C_j^0}N\ ; \ \ \ \ n_\ell (0) 
= e^{-\lambda_0}\frac{(\lambda_0)^\ell}{\ell!} \ ,
\eeq
where $\l_0$ is determined by 
$\sum_\ell \ell \ n_\ell(0) = \l_0 = \sum_j j
c_j(0)$.

It is easy to check that equations (\ref{LRc}) are solved by $c_j(\t) = c_j(0)\;
b(\t)^j$ provided $\frac\NN{b} \frac{d b}{d \t}= - 1$.  It is convenient to
introduce the {\it generating function} 
\begin{equation} \label{defG}
G(b) = \sum_{j=2}^k c_j(0)\; b^j \ .
\end{equation}
Derivative(s) of $G$ with respect to its argument will be denoted by prime(s). 
We have that $\NN(\t) = b(\t) G'(b(\t))$. In addition, we define $\g(\t) =
\sum_j c_j(\t)=G(b(\t))$. We deduce  the equation for $b(\t)$:
\beq
\label{tdib} \frac{d\g}{d\t} = \frac\NN{b} \frac{d b}{d \t}= - 1 \ \ 
\Rightarrow \ \ \t = \gamma(0) - \gamma(\t) = \sum_{j=2}^k
c_j(0) (1-b(\t)^j) \ .  
\eeq 
The interpretation of the equation above is just that at each step of the leaf
removal one equation is eliminated.

The solution to (\ref{LRn}) remains Poissonian at all times for all $\ell
\geq 2$. Substituting $n_\ell(\t) = e^{-\lambda(\t)}
\frac{\lambda(\t)^\ell}{\ell!}$ we obtain an equation for $\l(\t)$: 
\beq
\frac{d\lambda}{d\t} = - \frac{\sum_{j\geq2} j(j-1)c_j(\t)}{(\sum_{j\geq 2}
jc_j(\t))^2} \lambda(\t) =
-\left[\frac{G''(b)}{G'(b)^2}\right]_{b=b(\t)} \l(\t) \ , 
\eeq 
with the initial condition imposed by  $\l(0) = \l_0= \sum_j j c_j(0) = G'(1)$. 
From (\ref{tdib}) we get $\frac{d\t}{db} = -G'(b)$ so that 
\beq 
\frac{d\lambda}{db} =
\frac{d\lambda}{d\t}\frac{d\t}{db} = \frac{G''(b)}{G'(b)} \l \ , 
\eeq
which is solved by 
\beq \label{lam_b}
\lambda(b) = G'(b) \ , 
\eeq
where the normalization is fixed by the initial condition for $\lambda$.
(\ref{tdib}) and (\ref{lam_b}) determine $b(\t)$ and $\l(\t)$, which
describe the evolution of the formula under the action of the leaf removal
algorithm.

\subsection{Static Phase Transitions} \label{sec:32}

The structure of the subformula remaining at the end of the leaf-removal (if
any) is indicative of the nature of the phase corresponding to typical formulas,
uniformly drawn at fixed $\{C_j^0\}$. Three phases are possible: the {\it
unclustered} phase where formulas are satisfiable and the solutions form a
unique cluster; the {\it clustered} phase where solutions are divided into many
clusters; and the {\it unsat} phase where the typical formula is not satisfiable
\begin{enumerate}
\item {\it Clustering transition}:
The leaf removal algorithm starts from $b=1$, then $b$ decreases according to
(\ref{tdib}) and the
algorithm stops at the largest value of $b$ such that $n_1 = 0$, \ie there
are no more variables with unique occurrence. We have
\begin{eqnarray*}
  n_1 &=& \sum_{j=2}^k j c_j - \sum_{\ell>1} \ell n_\ell =  b G'(b) -
\sum_{\ell>1} \ell  e^{-\lambda(b)}             
        \frac{\lambda(b)^\ell}{\ell!} \\
    &=&  b  \lambda(b) -  e^{-\lambda(b)} \lambda(b) 
        \left[ e^{\lambda(b)} -1 \right]
    = \lambda(b) \left[ b - 1 + e^{-\lambda(b)} \right] \ ,
\end{eqnarray*}    
therefore 
\beq\label{conddinamica}
n_1 = 0 \ \ \ \Leftrightarrow \ \ \ 1-b = e^{-\lambda(b)} = e^{-G'(b)} \ .
\eeq 
This equation always has the solution $b=0$, that gives $c_j=0$ for
all $j$ when the algorithm stops. This corresponds to a backbone-free
formula whose solution space is connected. On the other hand, if this equation
admits non-trivial solutions $b> 0$, the algorithm stops when $b$ is equal to
the largest of them, \ie it is unable to eliminate all clauses in the formula.
Then the space is clustered and the largest
solution represents the fraction of variables in the 
backbone of each cluster~\cite{Mezard03, Cocco03}. 

In the pure $(k,d)$-UE-CSP case, \ie when $c^0_j = \a \d_{j,k}$,
the critical ratio at which clustering appears decreases with $k$,
from $\alpha_d(3)\simeq 0.818$ to $\alpha_d(k)\simeq \log k/k$ at large $k$.

\item {\it Sat/unsat transition}: The formula is satisfiable when the subformula
left by the removal algorithm has a solution. This happens with high probability
if and only if the number of equations, given by $G(b)$, is smaller than the
number of variables, $\sum_{\ell\geq 2} n_\ell$~\cite{Mezard03, Cocco03}.
Using the condition $n_1=0$, the satisfiability condition is
\beq\label{condsat}
G(b) \leq b+(1-b)\log(1-b) \ .
\eeq
For $(k,d)$-UE-CSP, the critical ratio at which formulas go from typically satisfiable to
typically unsatisfiable increases with $k$,
from $\alpha_s(3)\simeq 0.918$ to $\alpha_d(k) \to 1$ at large $k$.
\end{enumerate}

\subsection{The potential for the backbone}
\label{sec:surfaces}

The outcome of the previous section can be summarized as follows. We considered
a formula specified by a set $\{c^0_j\}_{j=2,\cdots,k}$, or equivalently by the
generating function (\ref{defG}). In the following we will drop the superscript
$0$ to simplify the notation. We define the {\it potential}
\beq\label{potedef}
V(b) = -G(b) +b +(1-b) \log(1-b) \ .
\eeq
The condition $n_1=0$ (\ref{conddinamica}), is equivalent to
$V'(b)=0$. Thus, if $V(b)$ has a single minimum in $b=0$, the solution space is
not clustered, while if there is another minimum at $b\neq 0$, there are
clusters. Moreover, the condition for satisfiability (\ref{condsat}), is
that at the secondary minimum $V(b) \geq 0$. Examples are given in
figure~\ref{fig_pot}.

The sat/unsat surface $\Si_s$, that separates the sat and the unsat phase, is
defined by the condition:
\beq\label{asdef}
\Si_s \equiv \{ c_j : V(b)=0  \ \mbox{and} \ V'(b)=0 \ \mbox{admit a solution} \
b>0 \} \ .
\eeq
The clustering surface $\Si_d$, that separates the
clustered and unclustered regions, is defined similarly by
\beq\label{dyna}
\Si_d \equiv \{ c_j : V'(b)=0  \ \mbox{and} \ V''(b)=0 \ \mbox{admit a solution}
\ b>0 \} \ .
\eeq
The equations above have to be interpreted as coupled equations for $(b,c_j)$;
therefore $\Si_{s},\Si_d$ have dimension $k-2$ and are surfaces in the
space
$\{c_j\}_{j=2,\cdots,k}$ of dimension $k-1$.
Note that in (\ref{asdef}) and (\ref{dyna}), one must always choose
the largest solution for $b$, to which we will refer as $b_s$ and $b_d$,
respectively. 

In addition to the previous sets, in the following a special role will be played
by the condition $2 c_2 = 1$, or equivalently $V''(0)=0$, 
that defines the {\it contradiction surface} $\Si_q$:
\beq\label{contra}
\Si_q \equiv \{ c_j : V''(0)=0 \} \ .
\eeq
The surface $\Si_q$ is simply a hyperplane of dimension $k-2$.

\subsection{The phase diagram}

\begin{figure}
\includegraphics[width=9cm]{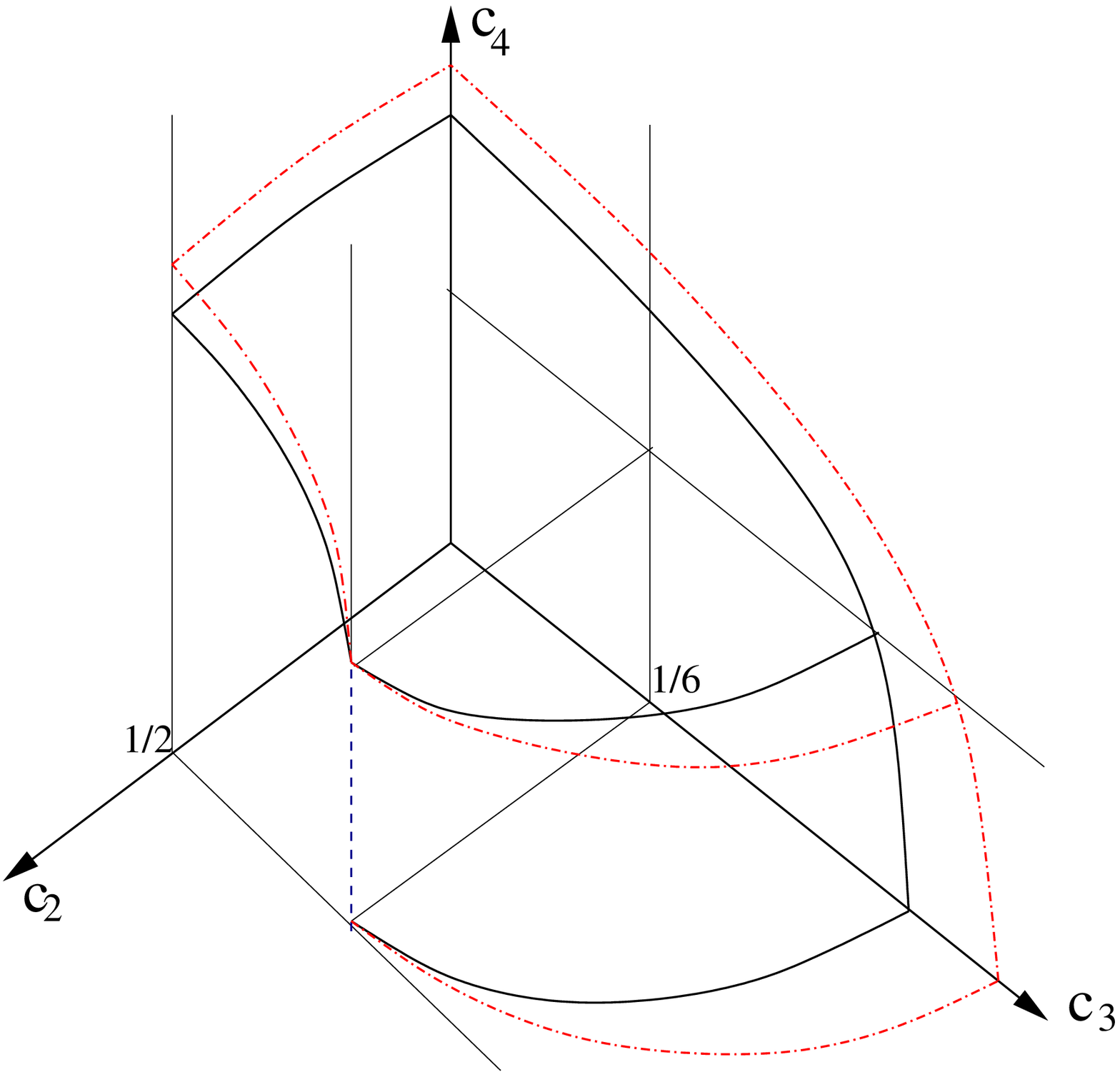}
\includegraphics[width=7cm]{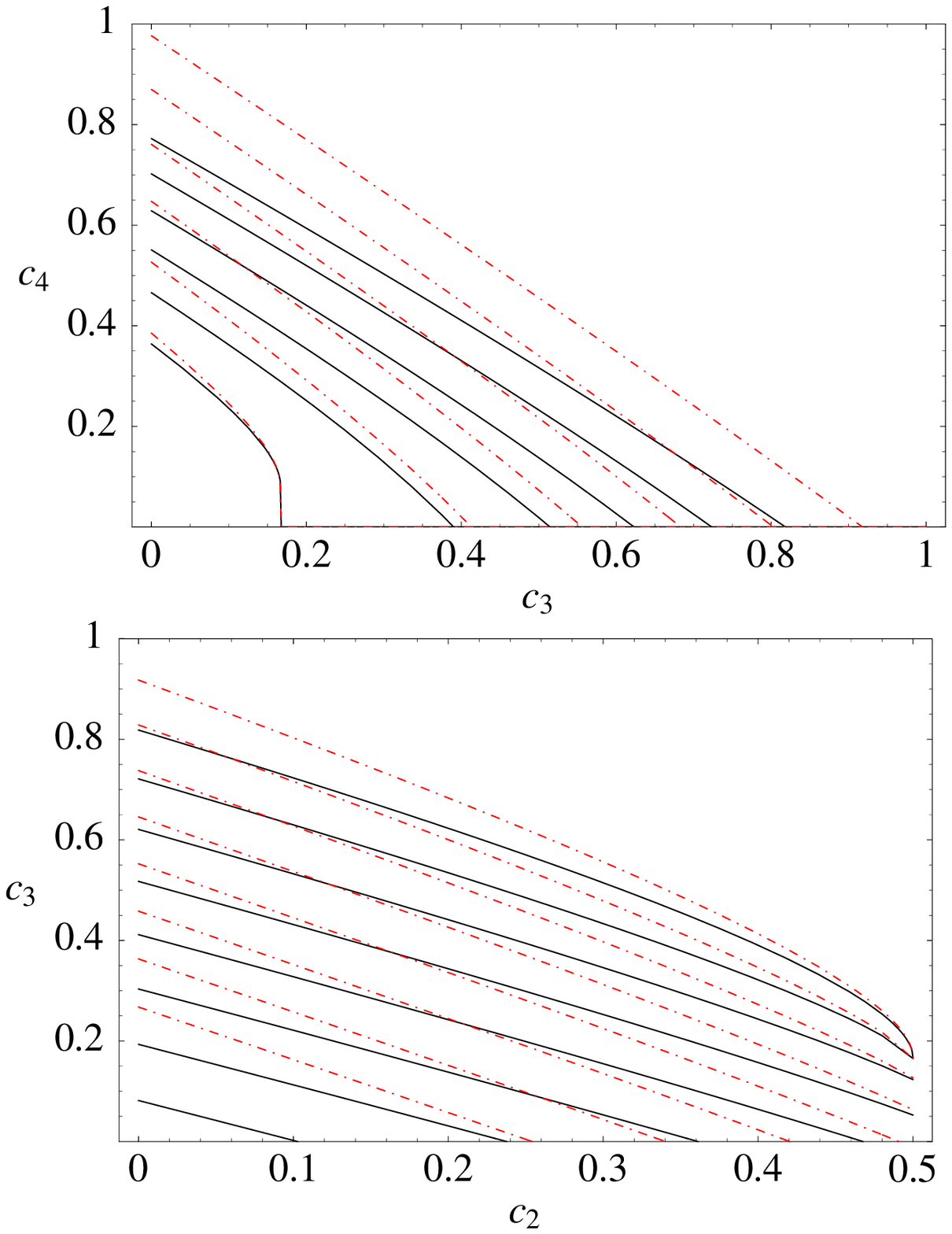}
\caption{{\bf (Left)} Schematic phase diagram of $k$$=$4-UE-CSP. The full (black)
curve is the surface $\Si_d$, the dot-dashed (red) surface is $\Si_s$. The two
surfaces meet along a portion of the line $\Si_{crit}$, defined by $c_2=1/2$ and
$c_3=1/6$ and represented as a dashed (blue) line. {\bf (Right, top and bottom)}
The sections of $\Si_d$ (full, black) and of $\Si_s$ (dot-dashed, red), at fixed
$c_2$ (=~0,~0.1, 0.2, 0.3, 0.4, 0.5 from top to bottom) as a function of $c_3$
on the top panel, and at fixed $c_4$ (=~0,~0.1, 0.2, 0.3, 0.4, 0.5, 0.6, 0.7
from top to bottom) as a function of $c_2$ in the bottom one. The lines
corresponding to $c_4=0$ also represent the phase diagram of 3-UE-CSP.}
\label{dia_fase_vero}
\end{figure}

We draw a phase diagram in the space of the $c_j$ by representing 
surfaces $\Si_{s},\Si_{d},\Si_{q}$. 
We focus on the region $c_j \in [0,1]$ for $j=3,\ldots,k$ and  
$c_2 \in [0,1/2]$. Indeed, if one of the $c_j >1$, the system is
surely in the unsat phase \cite{Connamacher04} while if $c_2 > 1/2$ the
algorithm
discussed above find a contradiction with very high probability.

Examples of the phase diagram are in figure~\ref{dia_fase_vero} for $k=3$ and
$k=4$. There are some special ``lines'' (\ie intersections of surfaces) 
on which we will concentrate. 
\begin{enumerate}
\item Recall that $\Si_q$ is defined by $V''(0)=0$ and
note that $V'(0)=0$ for all
$b$, $c_j$. Thus, on $\Si_q$, the point $b=0$ is a solution of
both equations (\ref{asdef}) and (\ref{dyna}).
The surfaces $\Si_{s},\Si_{d}$ are defined by the existence of
solutions with $b > 0$, but they might intersect $\Si_q$ if for some
values of $\{c_j\}$ the solution with $b>0$ merges with the solution $b=0$.
This happen when $V'''(0)=0$, as this is the limiting case in which a saddle 
at $b=b_d>0$ and a secondary minimum at $b=b_s>0$ can merge for $b_d,b_s\to 0$. 
The condition $V'''(0)=0$ is equivalent to $c_3 = 1/6$, and this defines the 
$k-3$-dimensional surface
\beq
\Si_{crit} \equiv \{ c_j : c_2= 1/2,c_3=1/6 \} \ ,
\eeq 
to which we will refer as
{\it critical surface}. It is easy to see that the three surfaces
$\Si_{s},\Si_d,\Si_q$ are
tangent to each other on the region of the critical surface where they
intersect. To show that
one must consider a displacement $c_3 = 1/6+\e$ and show that (\ref{dyna}),
(\ref{asdef}) admit a solution with $b_s,b_d \sim \e$ if $c_2 -1/2 \sim \e^2$.
We say that in this case the phase transitions are of {\it second order} because
the order parameter $b$ vanishes continuously at the transition.
\item There is no {\it a priori} reason for which the three surfaces must cross
at
$\Si_{crit}$. In fact, the solutions at $b>0$ might also disappear
discontinuously,
like in figure~\ref{fig_pot}, and the surfaces $\Si_s$ and $\Si_d$ can intersect
the
surface $\Si_q$ in regions different from $\Si_{crit}$.
This does not happen for $k=3$ but happens for $k=4$ for large $c_4$, see
figure~\ref{dia_fase_vero}.
In this case the transition is called {\it first order} because the order
parameter jumps at the transition.
\end{enumerate}
The generic phase diagram for all $k$ has the shape of the one for $k=4$ which
we report in figure~\ref{dia_fase_vero}, left panel.

\section{Search Trajectories in the Space of Formulas}
\label{sec:dyna}

The heuristics we defined in section~\ref{sec:def} enjoy the property that, 
after any number of steps of
the algorithm, the reduced formula is uniformly distributed over the set of
remaining $N-T$ variables conditioned to the numbers $C_j(T)$ of clauses of
length $j$ ($=2,...,k$) \cite{Chao90,Achlioptas01}. 
This statistical property, combined with
the concentration phenomenon taking place in the large $N$ limit, allows us to
study the evolution of the average clauses densities $c_j(t)=C_j(T)/N$ on the
time scale $t=T/N$ (fraction of assigned variables), which defines a {\it
trajectory} in the $c_j$'s space. Note that these $c_j(t)$ are defined with
respect to $N$, therefore the actual clause density for the reduced system of
$N-T$ variables are $\wt c_j(t) = c_j(t)/(1-t)$. {\it The trajectory of the $\wt
c_j(t)$ moves in the $c_j$ space of the previous section}\footnote{The reader
  should keep in mind this change of notation to avoid confusion in the
  following arguments}.

Initially we have $c_j(0)=\a\; \d_{jk}$, \ie the evolution starts on the $c_k$
axis at $c_k = \a$. The evolution equation for the densities take the form of
first order differential equations,
\beq\label{eq_gen_cont} 
\dot c_j = \frac{(j+1) c_{j+1} - jc_j}{1-t} -
\r_j(t) \ .  
\eeq 
The interpretation of the equations above is the following.  Let us consider an
interval $[t,t+dt]$ of continuous time that corresponds to $\D T\sim N dt$ time
steps of the algorithm. The first term on the r.h.s. arises from the decrease by
one of the length of the clauses that contained the variable just assigned by
the algorithms during this interval. The second term corresponds to an
additional loss of clauses which is present when the variable is selected from a
clause of length $j$: as the heuristics explicitly chooses an equation (and a
variable therein) of length $j$ with probability $p_j$ (see
section~\ref{sec:def}), 
this equation will be reduced irrespectively of the number of
other occurrences of the variable. Hence $\rho_j(t)$ is given, for $j \geq 1$,
by
\beq\label{rhoj}
\rho_j(t) = \lim _{\Delta T\to\infty} \lim _{N\to\infty} \frac
1{\Delta T} \sum _{T=tN }^{tN+\Delta T-1} \left( p_{j} - p_{j+1}
\right) 
\equiv \la p_{j} - p_{j+1} \ra
\ ,
\eeq
where both $p_j,p_{j+1}$ depend on their arguments (numbers of clauses)
and $\la \bullet\ra$ represents the average over $\D T$ defined in (\ref{rhoj}).
Here $p_{k+1}\equiv 0$.
Note that the case $j=1$ is special as all clauses of length one that are
produced are
immediately eliminated. On average
\beq
\r_1 \equiv \frac{2 c_2}{1-t}
\eeq
clauses of length 2 become of length 1 and are then eliminated by unit 
propagation. The total fraction of eliminated clauses is
\beq\label{cond_rho_1}
\dot\g(t) \equiv -\sum_{j=2}^k \dot c_j(t) = \frac{2 c_2(t)}{1-t} + \sum_{j=2}^k
\r_j(t)
= \sum_{j=1}^k \r_j(t) \leq 1
 \ ,
\eeq
where the last inequality follows from (\ref{rhoj}).
As only clauses of length one are eliminated, the violation of (\ref{cond_rho_1}) 
can only happen if too many such clauses are generated. 
This corresponds to $\rho_1 \to 1^-$;
in this case a contradiction occurs 
with high probability and the algorithm stops with the `Don't know' output.
When $\r_1 \to 1^-$, the
algorithm makes only unit propagations and $\r_j \to 0^+$ for all $j \geq 2$. 
For this reason we called the plane $\r_1 = 1$, \ie $\wt c_2 = 1/2$, {\it contradiction
surface}.

\subsection{Unit Clause (UC)}

In the UC heuristic variables are chosen at random when there is no unit 
clause. Hence $\rho _j =0$ for $j=2,\cdots,k$. The solution to
(\ref{eq_gen_cont}) is $c_j(t)=\a {k \choose j} (1-t)^j t^{k-j}$. The
algorithm will generate a contradiction with high probability (w.h.p.) if the
average number of unit clauses starts to build-up, i.e. if $2 c_2(t)/(1-t) \geq
1$. This gives an equation for the value of $\a$ at which the probability that
the algorithm finds a solution vanishes: for $k=3$, $\a_a^\mathrm{(UC)} = 2/3$.

\subsection{Generalized Unit Clause (GUC)}\label{sec-guc}

In the GUC heuristic the algorithm always fixes a variable appearing in the
shortest clauses. In the continuum limit $c_j = 0$ for $j$ smaller than a given
value; therefore we define
\begin{equation}
j^*(t) = \min \{ j : c_j(t) >0 \} \ ,
\end{equation}
the minimal length of clauses with positive densities. We also define
\begin{equation}
t^*(j) = \min [ t : c_{j-1}(t) >0]
\end{equation}
the time at which $j^*$ jumps down from $j$ to $j-1$. Essentially, the algorithm
picks one clause of length $j^*$ and assigns successively all the variables in
this clause until the clause disappears. But in doing so, other clauses of
length $j < j^*$ are generated and have to be eliminated to recover the
situation in which $C_j=0$ for all $j < j^*$; for this reason
$\r_{j^*}$ is not given exactly by $1/j^*$. 
When the number of generated clauses is so high that the algorithm is unable to
remove them, $c_{j^*-1}$ becomes different from 0 and $j^*$ jumps down by 1.
The resulting motion equations for the clause densities are, for $j \ge j^*(t)$:
\begin{equation} \label{eqmot}
\dot c_j (t) = \frac{ (j+1) c_{j+1} (t) - j c_j(t)}{1-t} 
- \delta _{j,j^*(t)} \left( \frac 1j - \frac{(j-1) c_j(t)}{1-t}\right) \ . 
\end{equation}
The transition times $t^*$ are given by  
\begin{equation} \label{cond}
\frac {c_j( t^* (j))}{1-t} = \frac 1{j(j-1)} \ ,
\end{equation}
where the algorithm is no more able to remove the clauses of length $j^*$
because too many clauses of length $j^*-1$ are being generated by propagations.

Comparing with (\ref{eq_gen_cont}) above, we observe that in the interval $t
\in [t^*(j+1),t^*(j)]$, where $j^*=j$, only two $\r_j$ are different from 0:
\beq \label{rho_j_GUC}
\r_{j^*} = \frac 1{j^*} - \frac{(j^*-1) c_{j^*}(t)}{1-t} \ , \ \ \ \ \r_{j^*-1}
= \frac{j^* c_{j^*}(t)}{1-t} \ ,
\eeq
the first representing clauses of length $j^*$ which are directly eliminated,
the second representing the clauses of length $j^*-1$ that are produced and
subsequently eliminated in the process. In this interval of time, the ratio
$c_{j^*}(t)/(1-t)$ increases from 0 to $1/j^*/(j^*-1)$ from condition
(\ref{cond}). Then
\begin{equation} \label{bounddotg}
\frac 1{j^*(t)} \le \dot\g(t)= ( \r_{j^*} + \r_{j^*-1}) \le  \frac 1{j^*(t)-1} \
,
\end{equation}
which is consistent with (but stronger than) (\ref{cond_rho_1}) above.

\section{Analysis of the ``dynamic'' phase diagram}
\label{sec:cinq}

\begin{figure}
\centering
\includegraphics[width=10cm]{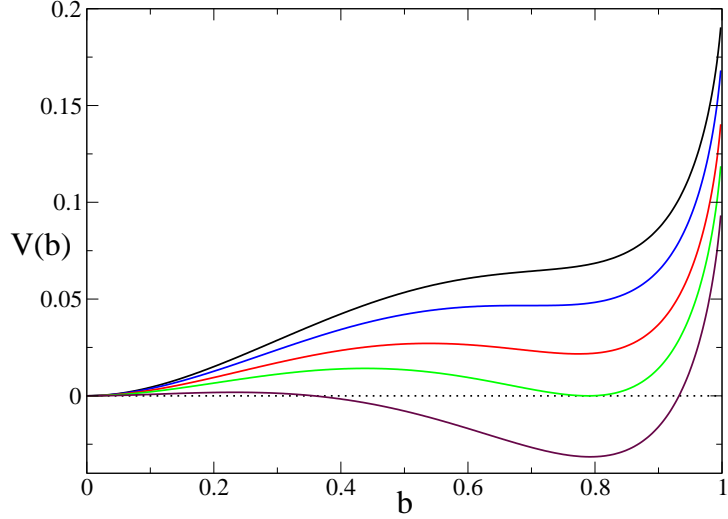}
\caption{An example of the potential $V(b;t,\a)$ plotted (from top to bottom) at
times $t=\{0,t_d=0.02957,0.07327,t_s=0.11697,0.20642\}$ during the evolution of
a $(3,d)$-UE-CSP formula with $\a = 0.8$ under the UC heuristic. In the
unclustered region it is a convex function of $b$ with a global minimum in
$b=0$. On the clustering line $t_d$ it first develops a secondary minimum. On
the sat/unsat line the value of $V$ at the secondary minimum becomes equal to 0.
}
\label{fig_pot}
\end{figure}

Consider now a given heuristic, and a generic $(k,d)$-UE-CSP formula specified
by its clause-to-variable ratio $\a$. The formula, in the $c_j$ space,
starts on the axis $c_k$ at $c_k = \a$. The evolution of the formula under the
action of the algorithm is represented by a {\it trajectory}
$\{c_j(t,\a)\}_{j=2,\cdots,k}$ or equivalently by $G(b;t,\a) = \sum_{j=2}^k b^j
c_j(t,\a)$, that depends on $\a$ through the initial condition $G(b;0,\a) = \a
b^k$. We define a potential $V(b;t,\a)$ by replacing in (\ref{potedef})
$G(b) \to G(b;t,\a)/(1-t)$; the normalization $(1-t)$ is due to the fact that
the $c_j = C_j/N$ are divided by $N$ instead of $N-T$.

We follow the evolution of the formula by looking at the times at which the
trajectory starting at $c_k = \a$ at time $0$ 
crosses the surfaces $\Si_{s},\Si_d,\Si_q$ defined in
section~\ref{sec:surfaces}, 
which we call $t_s(\a),t_d(\a),t_q(\a)$ respectively. 
As an example, in figure~\ref{fig_pot} we report the potential at different
times 
during the evolution of a formula according to the UC heuristic for $\a >
\a_a^{(UC)}$.

\begin{figure}
\includegraphics[width=8cm]{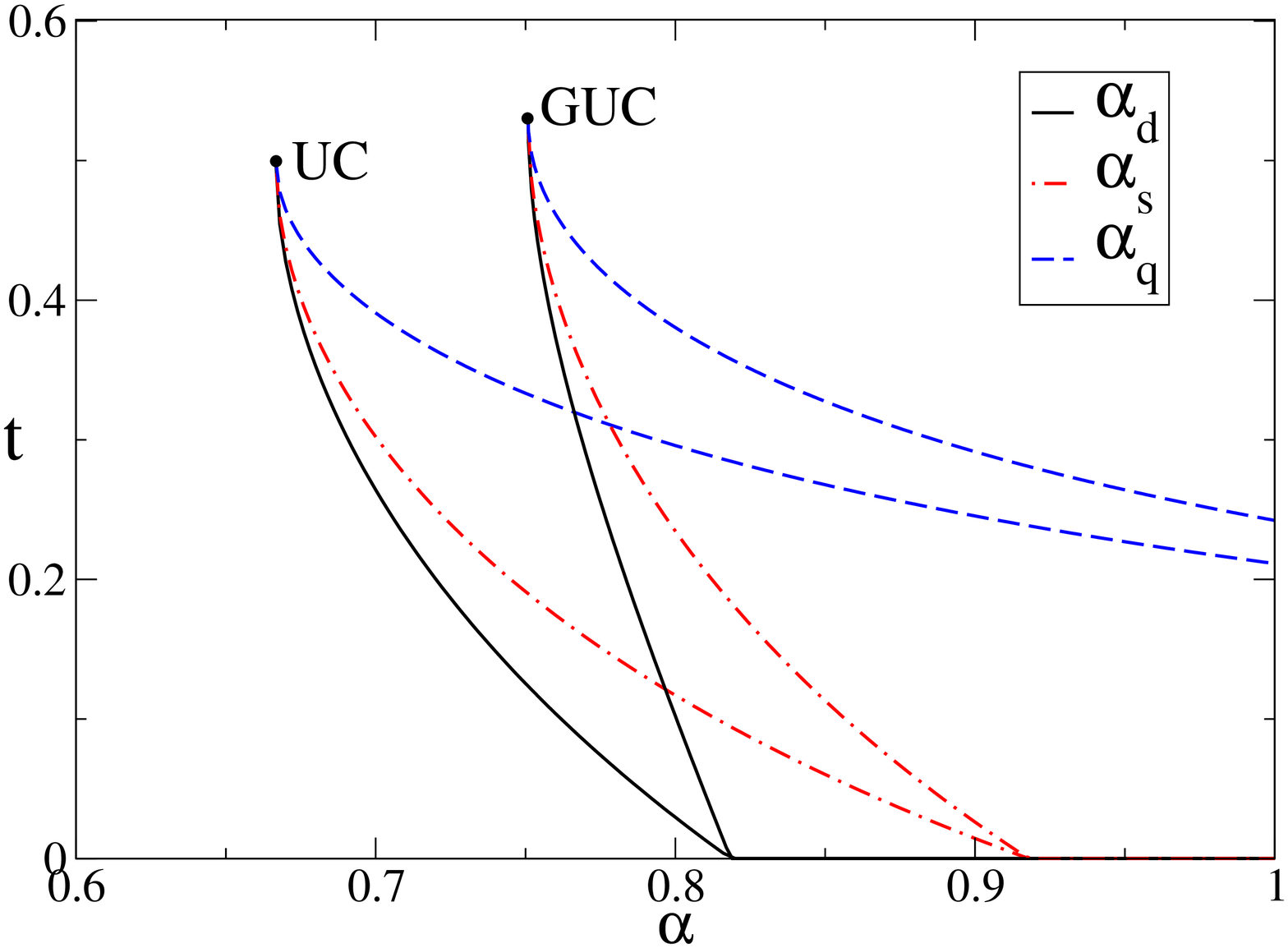}
\includegraphics[width=7.5cm]{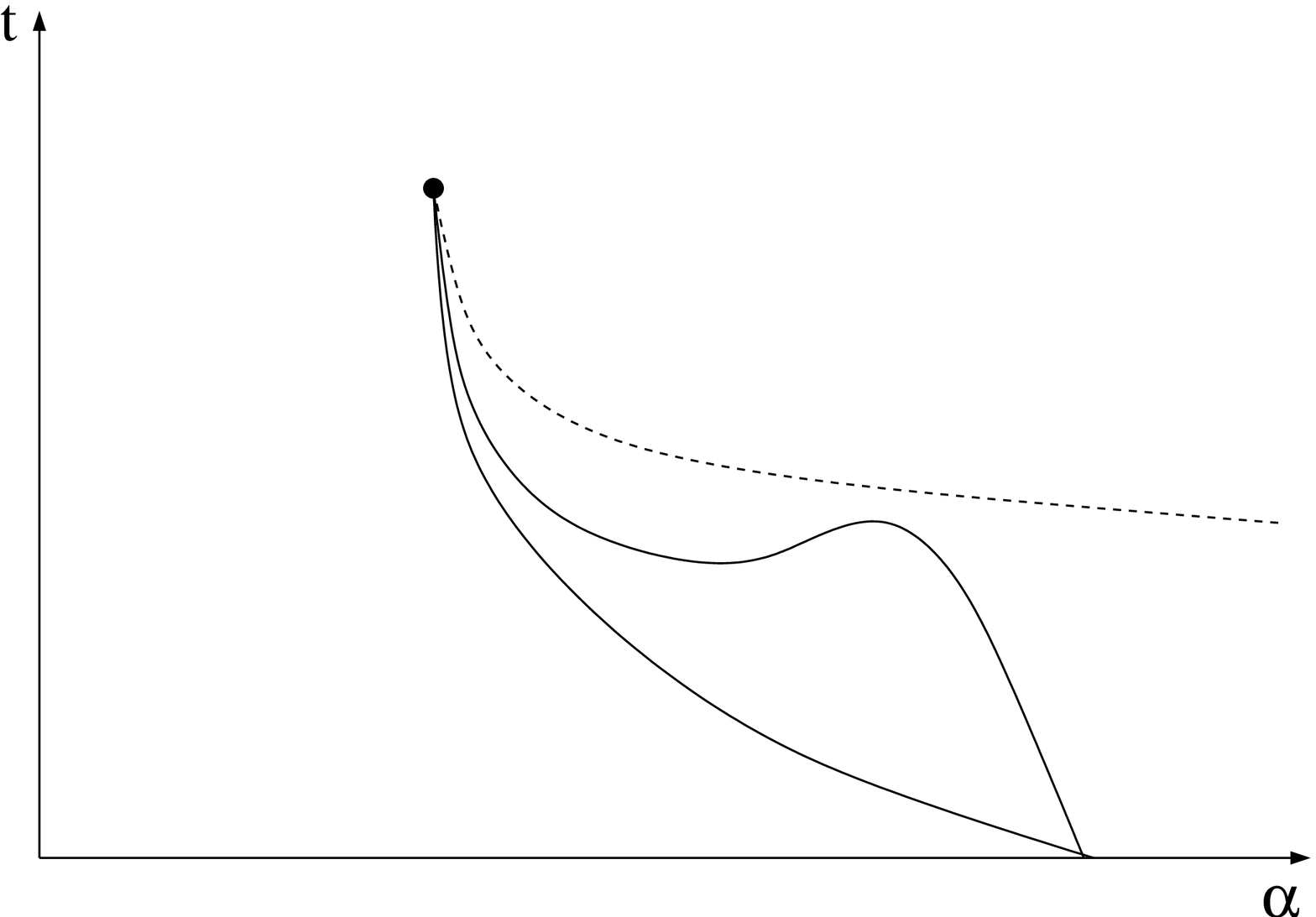}
\caption{{\bf (Left)} Phase boundary lines in the $(t,\a)$ plane for the UC and
GUC heuristics for $k=3$. The three lines meet at the critical point
$(t_a,\a_a)$ at which the algorithm is no more able to find a solution (black
dot). 
{\bf (Right)} The generic shape of the clustering and of the sat/unsat lines.
The possibility of a maximum cannot be excluded, but in any case $t$ must be a
single-valued function of $\a$, meaning that if the algorithm enters the cluster
(or unsat) phase it cannot escape at later times.
}
\label{dia_fase}
\end{figure}

We draw a ``dynamic phase diagram'' by representing in the $(t,\a)$ plane
the lines separating the unclustered, clustered, unsat and
contradiction phases, which we call $\a_d(t),\a_s(t),\a_q(t)$ and are just the
inverse of the times defined above. Examples in the case of the UC and GUC
heuristics are given in figure~\ref{dia_fase}.

From the general properties of the function $V(b;t,\a)$ we can deduce a number
of properties of the lines $\a_d(t),\a_s(t),\a_q(t)$. We will show that
the three lines intersect at
a ``critical point'' $(t_a,\a_a)$, located at  $\a_a\leq\a_d$, 
under the more general conditions. This
implies that the algorithm stops working at the value $\a_a\leq \a_d$,
which is our central result: {\it Poissonian search algorithm cannot
find a solution in polynomial time in the clustered region}.

\subsection{Equations for the transition lines}

The generating function $G(b;t,\a)$ satisfy an evolution equation which is
easily derived from (\ref{eq_gen_cont}):
\begin{eqnarray}\label{eq_gen_G}
\dot G(b;t,\a) &=& \frac{1-b}{1-t} G'(b;t,\a) - F(b;t,\a) \ , \\ F(b;t,\a)
&\equiv& \frac{2 c_2(t)}{1-t} b + \sum_{j=2}^k \r_j(t) b^j =
\sum_{j=1}^k \r_j(t) b^j \ .
\end{eqnarray}

Performing the total derivative with respect to $t$ of the first condition
($V'=0$)
in (\ref{dyna}) for $(\a_d,b_d)$ and using the second condition, $V''=0$, we
have
\beq
\frac{\partial V'}{\partial \a} \frac{d \a_d}{dt} + \dot V' = 0 \ \ \ \
\Rightarrow
\ \ \ \
\frac{d \a_d}{dt} = - \frac{\dot V'(b_d;t,\a_d)}{\frac{\partial V'}{\partial
\a}(b_d;t,\a_d)} \ .
\eeq
Using the definition (\ref{potedef}) we have
\begin{eqnarray}
\dot V'(b;t,\a) &=& -\frac{1}{1-t} \left[ \dot G'(b;t,\a) +
\frac{G'(b;t,\a)}{1-t}
\right] \ , \\
\frac{\partial V'}{\partial \a} &=& -\frac1{1-t} \frac{\partial G'}{\partial \a}
=-\frac1{1-t} \sum_{j\geq 2} j b^{j-1} \frac{\partial c_j(t,\a)}{\partial \a} \
.
\end{eqnarray}
Then
\beq
\frac{d \a_d}{dt} =- \left. \frac{ \dot G'(b;t,\a) + \frac{G'(b;t,\a)}{1-t} }
{ \partial_\a G'(b;t,\a) } \right|_{\a=\a_d(t),b=b_d(t)} \ .
\eeq
Using (\ref{eq_gen_G}) and
differentiating with respect to $b$ we have
\beq
\dot G'(b;t,\a) + \frac{G'(b;t,\a)}{1-t} =  \frac{1-b}{1-t} G''(b;t,\a) -
F'(b;t,\a) \ .
\eeq
Now using 
$V''(b;t,\a) = - \frac{ G''(b;t,\a)}{1-t} + \frac1{1-b}$
and $V''(b_d,t)=0$ we have $\frac{ 1-b}{1-t} G''(b;t,\a) = 1$ for $b=b_d$ and
finally we get
\beq\label{dneg}
\frac{d \a_d}{dt} = -\left. \frac{ 1 - F'(b;t,\a)}
{ \partial_\a G'(b;t,\a) }\right|_{\a=\a_d(t),b=b_d(t)} \ .
\eeq
A very similar reasoning leads to the following equation for the sat/unsat line:
\beq\label{sneg}
\frac{d \a_s}{dt} = - \left. \frac{b - F(b;t,\a)}{\partial_\a G(b;t,\a)} 
\right|_{\a=\a_s(t),b=b_s(t)}
\ .
\eeq
The equation for the contradiction line is easily derived from its definition
$\wt c_2(t,\a) = \frac{c_2(t,\a)}{1-t} = \frac12$, which immediately gives
\beq\label{qneg}
\frac{d\a_q}{dt} = - 
\left. \frac{1 + 2 \dot c_2(t,\a)}{2 \partial_\a c_2(t,\a)} \right|_{\a=\a_q(t)}
 \ .
\eeq

\subsection{General properties of the transition lines}

We wish to show that the transition lines $t_d(\a)$,$t_s(\a)$ and $t_q(\a)$ in the $(\a,t)$ plane
are single-valued
functions of $\a$, and that they meet in a point $(\a_a,t_a)$ where they have infinite slope and are
therefore tangent to each other; the value $\a_a$ correspond
to a trajectory which is tangent to the crytical surface $\Si_{crit}$.

Our argument 
goes as follows:
\begin{enumerate}
\item
We defined $\a_a$ as the value of $\a$ for which the probability of finding a 
solution for the chosen heuristic vanishes.
Then the trajectory\footnote{Recall that we are here talking about average
trajectories.} corresponding to any $\a > \a_a$ must cross the contradiction
surface, while the trajectory corresponding to any $\a < \a_a$ must not cross
it, so that the trajectory corresponding to $\a_a$ must be \emph{tangent} to the
contradiction surface $\Si_q$. 
The latter trajectory is tangent to $\Sigma_q$ when
$\wt c_2(t) = 1/2$, $\frac{d}{dt} \wt c_2(t) = 0$; the solution to these conditions
gives $t_a$ and $\a_a$. \\
Moreover, $\wt c_2(t) = 1/2$ implies
that $\r_1 = 1$ which then implies $\r_j = 0$ for all $j\geq 2$, as already discussed.
Then we have
\begin{equation}\label{contra1}
\frac{d}{dt} \wt c_2(t) = 
\frac d {dt} \frac {2 c_2(t)}{1-t} = \frac{2 \dot c_2(t)}{1-t} + \frac{2 c_2(t)}{(1-t)^2} = 0
\ \ \ \Rightarrow \ \ \ \dot c_2(t) = -\frac{ c_2(t)}{1-t} = -\frac12 \ ,
\end{equation}
which, together with the equations of motion (\ref{eq_gen_cont}) and $\r_2 = 0$ gives
\begin{equation}\label{contra2}
-\frac{ c_2(t)}{1-t} = \frac{d c_2(t)}{dt} = \frac{3 c_3(t) - 2 c_2(t)}{1-t} 
\ \ \ 
\Rightarrow
\ \ \
\wt c_3(t) = \frac{c_3(t)}{1-t} = \frac13 \frac{ c_2(t)}{1-t} = \frac16 \ .
\end{equation}
Therefore the point where the trajectory for $\a=\a_a$ is tangent to the contradiction
surface belongs to the critical surface $\Si_{crit}$.
From equation~(\ref{qneg}) it is clear that since $\dot c_2 = -1/2$, the function $t_q(\a)$ has
infinite slope in $(t_a,\a_a)$, as in figure~\ref{dia_fase}.
\item
Next we show that the numerators of the fractions
appearing in $\dot \a_d(t)$ and $\dot \a_s(t)$ are strictly positive if $t <
t_q(\a)$, \ie in before a contradiction is found.
Using the definition (\ref{rhoj}) we can write:
\begin{eqnarray}\label{Fprimapp}
 F(b;t,\alpha) &=& \sum_{j=1}^k \rho_j(t) b^j = b \left[ \la p_1 \ra + \sum_{j=2}^k 
b^{j-2} (b-1) \left\langle p_j\right\rangle \right] \ , \\
 F'(b;t,\alpha) &=& \sum_{j=1}^k j \rho_j(t) b^{j-1} = \left\langle p_1 \right\rangle + 
\sum_{j=2}^k b^{j-2} \left[ 1 - j(1-b) \right] \left\langle p_j \right\rangle \, . \nonumber
\end{eqnarray}
The coefficients in front of $\left\langle p_j \right\rangle \geq 0$ 
in the sums above are always smaller than 1, independently of $j$, so that
\begin{eqnarray}\label{bound}
 F(b;t,\alpha) &\leq& b \left[ \left\langle p_1 \right\rangle + \sum_{j=2}^k \left\langle p_j \right\rangle\right] \leq b \ , \\
F'(b;t,\alpha) &\leq& \left\langle p_1 \right\rangle + \sum_{j=2}^k \left\langle p_j \right\rangle \leq 1 \,. \\
\nonumber
\end{eqnarray}
The functions $F(b;t,\a)$ and $F'(b;t,\a)$ are to be computed in $b=b_s(t,\a)$ or
$b=b_d(t,\a)$ in equations~(\ref{dneg}) and (\ref{sneg}). Both $b_s$ and $b_d$ are strictly
smaller than $1$ for all $(t,\a)$, as one can directly show from their definitions because
$V'(b\to 1) \to \io$.
Then the coefficients in the sums in (\ref{Fprimapp}) are strictly smaller than $1$, 
and the only
solution to $F=b$ or $F' = 1$ is $\la p_j \ra = \d_{1j}$, which happens only on the contradiction
line.
\item
The denominators in equations~(\ref{dneg}), (\ref{sneg}) 
are surely positive at $t=0$, as $G(b;0,\a) = \a b^k$
independently of the heuristic. If they remain positive
at all times, then $\dot \a_d(t),\dot \a_s(t) \leq 0$ at all times, or
equivalently $\frac{dt_d}{d\a},\frac{dt_s}{d\a} \leq 0$ at all $\a$, so that
$t_d,t_s$ always increase on decreasing $\a$. \\
The other possibility is that the denominator in (\ref{dneg}) crosses zero 
and become negative, 
leading to a maximum in $t_d(\a)$, which will then decrease on decreasing $\a$. 
Possibly the denominators can vanish again, giving rise to a sequence of maxima
and minima, see right panel of figure~\ref{dia_fase}. \\
What is important is that the numerator is always strictly positive, and as a
consequence $t_d(\a)$ or $t_s(\a)$ are single-valued functions of $\a$. In fact,
for $t_d(\a)$ or $t_s(\a)$ to be multiple-valued functions of $\a$, at some point
their slope must become infinite, which is excluded by the analysis above.
\item
The statement above, that $t_d(\a)$ and $t_s(\a)$ are
single valued functions of $\a$, implies that {\it if a trajectory enters the
clustered or unsat phase, it cannot exit from it}. This is enough to show that
$\a_a \leq \a_d$; in fact, the trajectory for $\a=\a_a$ cannot start inside the clustered
phase, as it would not be able to escape and reach the origin, which is required
to find a solution.
\item
In general the function $\wt c_2(t)$ increases until it reaches a maximum
and then decreases to 0. For $\a=\a_a$ the value at the maximum is $\wt c_2 =
1/2$. For $\a > \a_a$, the value at the maximum is $\wt c_2 > 1/2$, therefore
the contradiction $\wt c_2 = 1/2$ is reached before the maximum, when $\wt c_2$ is still
increasing. Then $\frac{d}{dt} \wt c_2 > 0$ at the contradiction point.
Performing a simple computation similar to equations~(\ref{contra1}),
(\ref{contra2}), one can show that the trajectories for $\a > \a_a$ meet the
contradiction surface at $\wt c_3 > 1/6$.
Notice then that, as it is evident in figure~\ref{dia_fase_vero}, the
trajectories corresponding to $\a > \a_a$ must enter first the clustered and
then the unsat phases in order to reach the contradiction surface, therefore
for $\a < \a_a$ one has $t_d(\a)<t_s(\a)<t_q(\a)$.
On the contrary the
trajectories corresponding to $\a < \a_a$ must stay away from the clustering
and sat/unsat surfaces, otherwise
they could not exit and should meet the contradiction surface: therefore
for any $\a < \a_a$, $t_d(\a)$ and $t_s(\a)$ do not exist. 
For $\a \to \a_a^+$, as the surfaces $\Si_d,\Si_s,\Si_q$ are tangent in $\Si_{crit}$,
one has $t_d(\a_a)=t_s(\a_a)=t_q(\a_a)=t_a$ and the three curves have infinite
slope as all the numerators in equations~(\ref{dneg}), (\ref{sneg}), (\ref{qneg})
vanish on the contradiction surface.
This is indeed what
is observed in figure~\ref{dia_fase} for the UC and GUC heuristics, and this
argument confirms that this is the generic behavior for all the heuristics in
the class considered here.
\end{enumerate}

This structure is particularly evident for UC, where 
\beq\label{G_UC} G^\mathrm{(UC)}(b;t,\a)= \a [1-(1-b)(1-t)]^k - \a t^{k-1} [ kb
(1-t) + t ] \ .
\eeq
From (\ref{G_UC}) it is straightforward to check that $\partial_\a
G(b;t,\a) > 0$, $\partial_\a G'(b;t,\a) > 0$, if $b > 0$. Then, as $F(b;t,\a) =
\frac{2 b
c_2(t)}{1-t}$ for UC, both $\dot \a_d(t)$ and $\dot \a_s(t)$ are proportional to
$ \frac{2 c_2(t)}{1-t} -1$. This means that $\a_s$,$\a_d$ are decreasing
functions of $t$ below the contradiction line. 

The conclusion is that for a generic Poissonian heuristic, the three lines
cross at a critical point $(t_a,\a_a)$ which depends on the heuristic. Above
$\a_a$ the heuristic will cross all the lines and find a contradiction. From the
properties of the dynamical line, we have that generically $\a_a \leq \a_d$,
that is {\it no Poissonian search heuristic can find a solution in polynomial
time
above $\a_d$}, as stated at the beginning of this section. The natural question
is then if there exists an heuristic that saturates the bound, \ie such that
$\a_a = \a_d$. From the discussion above it is clear that this is possible only
if $\dot\a_d(t) \equiv 0$, \ie the dynamical line in the $(t,\a)$ plane is a
straight vertical line, which is possible only if the numerator in (\ref{dneg})
is identically vanishing.

\subsection{Optimality of GUC}

It is quite easy to see that GUC is the heuristic that {\it locally} optimizes
the numerator in (\ref{dneg}). Indeed, from the definition
$F'(b;t,\a)=\sum_{j=1}^k j b^{j-1} \r_j$  and the bound $F'(1,t) \leq 1$, it is
clear that $F'(b;t,\a)$ is maximized by maximizing $\r_j$ for the smallest
possible $j$, \ie by picking clauses from the shortest possible ones, that is
GUC. Unfortunately a general proof of the optimality of GUC for finite $k$ seems
difficult, because one should prove that GUC optimizes globally the clustering
line, and also control the denominator in (\ref{dneg}). In this section we will
show that for $k\to\io$, GUC is optimal in the sense that $\dot\a_d \equiv 0$
and $\a_d=\a_a$ at leading order in $k$.

\begin{figure}
 \includegraphics[width=0.5\textwidth]{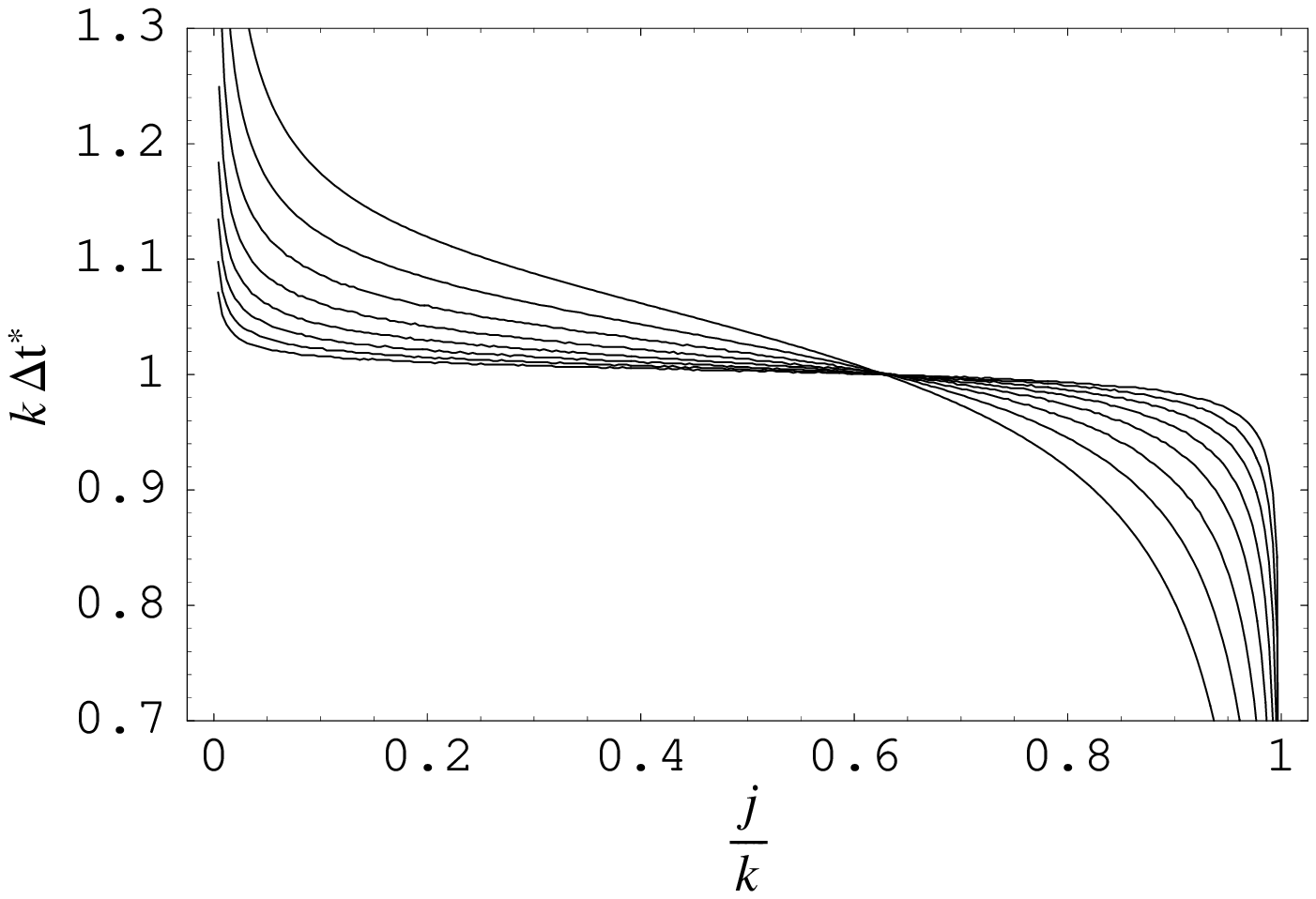}
 \includegraphics[width=0.5\textwidth]{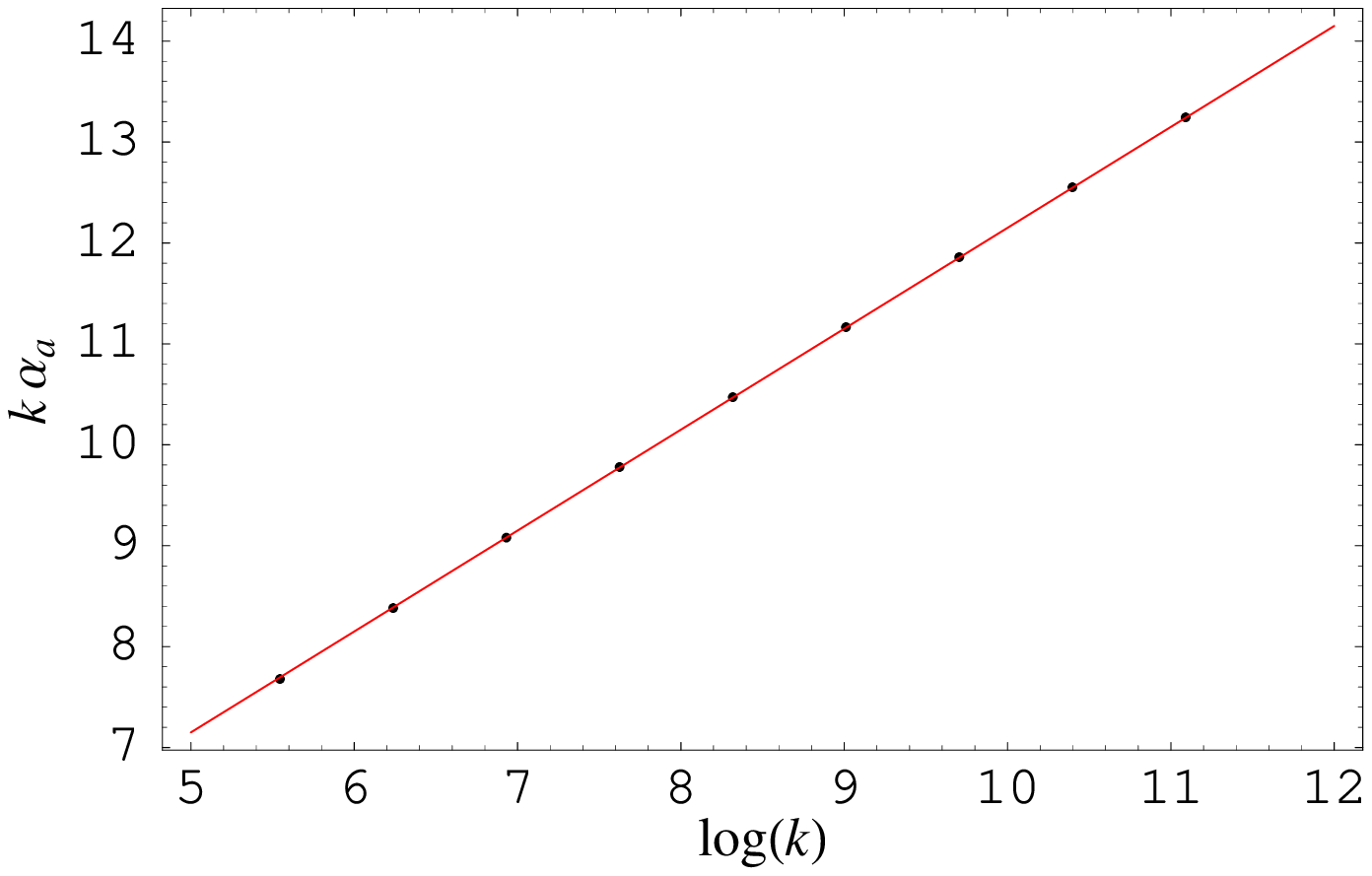} \\
 \includegraphics[width=0.5\textwidth]{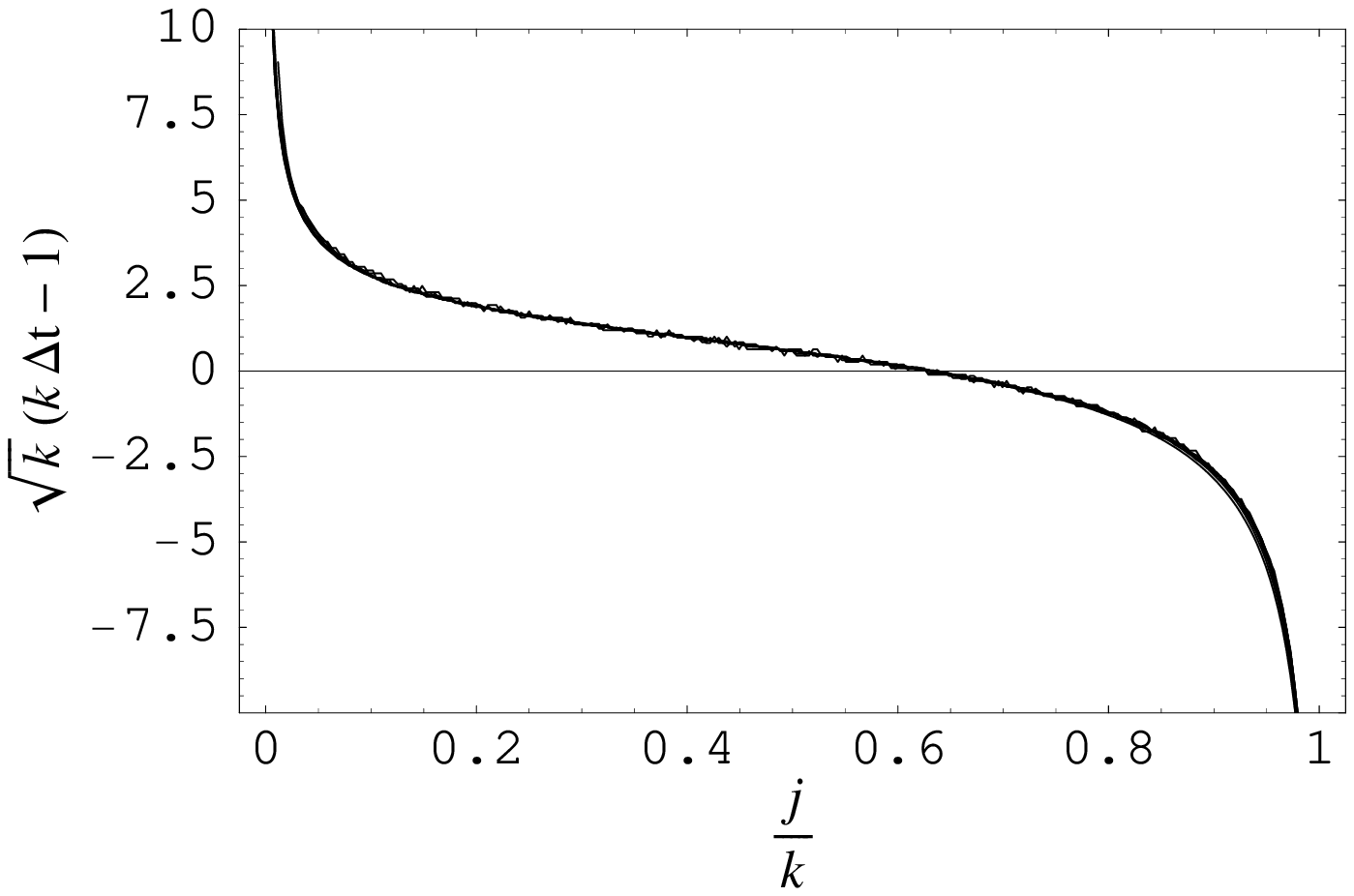}
 \includegraphics[width=0.5\textwidth]{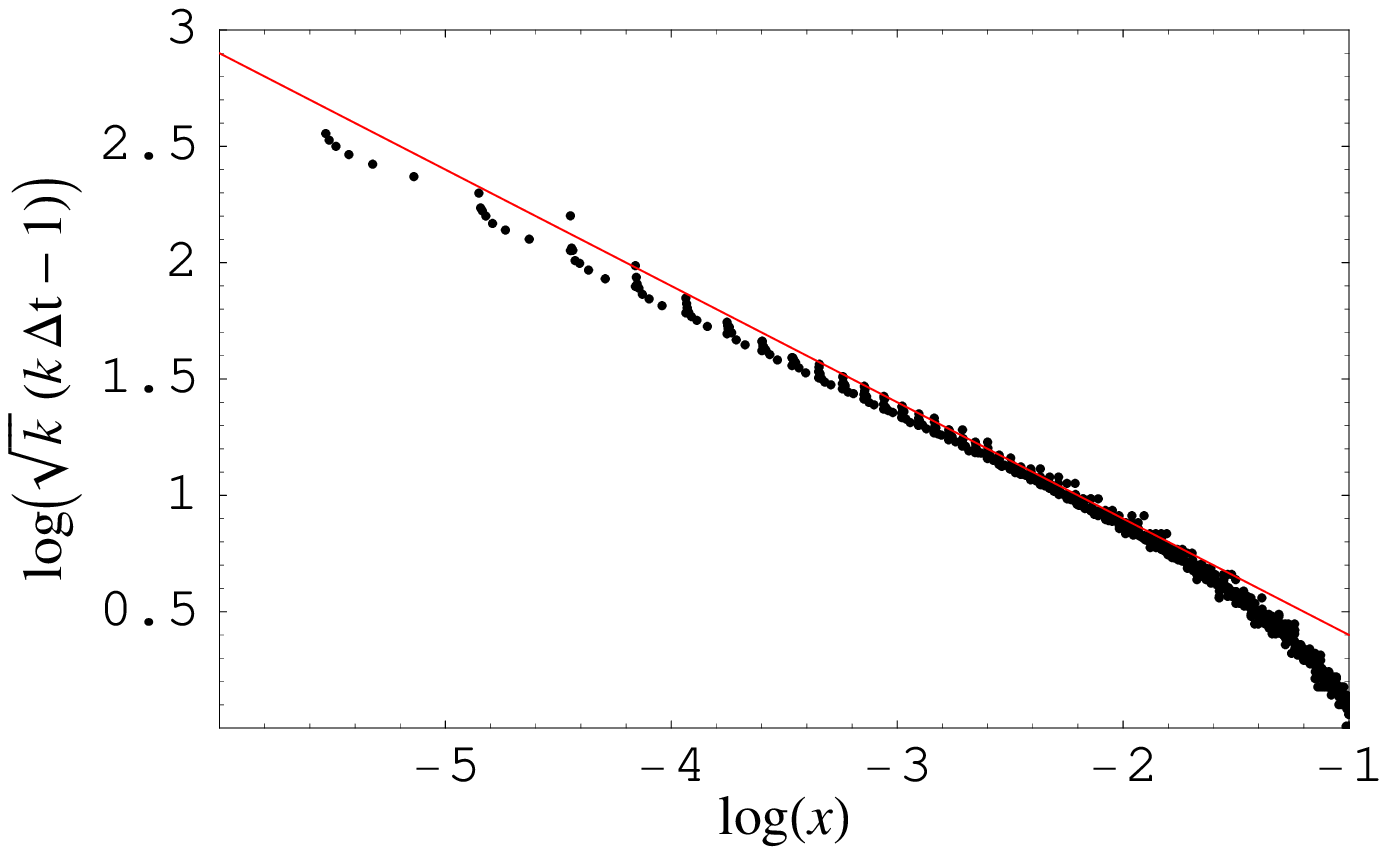}
 \caption{
  Finite size scaling results for GUC at large $k$. 
  \emph{Top~Left}~Each curve shows the values of $k[t^*(j)-t^*(j+1)]$ as a function of $j/k$ for $k=2^8,2^9,\dots,2^{16}$ (from the farthest to the closest curve to 1), and was obtained by integrating the equations of motion (\ref{eqmot}) by finite differences. For each $k$, the value of $\alpha$ used is $\alpha_a^\mathrm{GUC}(k)$, determined as the value of $\alpha$ for which the maximum reached by $2 c_2(t) / (1-t)$ is 1. 
  \emph{Top~Right}~Data points of $\alpha_a^\mathrm{GUC}(k)$ versus $\log k / k + 2.15 / k$ (full red line).
  \emph{Bottom~left}~The same data as above, plotted as $\{k\times[t^*(j)-t^*(j+1)]\}\times k^{1/2}$. The curves ``collapse'', showing $f(x)$ and confirming the value of $\nu = 1/2$. 
  \emph{Bottom~right}~By plotting the same curves on logarithmic scale it is easily seen that for $x$ close to 0 $f(x) \simeq x^{-\mu}$ with $\mu = 1/2$, corresponding to the slope of the full red line.
 }
 \label{fig_tstar}
\end{figure}

From the definition $\gamma (t) =- \sum _{j=2} ^k c_j(t)$ and integrating over
time the bound
(\ref{bounddotg}), we have for GUC:
\begin{equation} 
\alpha - \int  _0 ^t \frac {dt'}{j^*(t')-1} 
\le  -\gamma (t) \le \alpha - \int _0 ^t  \frac {dt'}{j^*(t')} \ . 
\end{equation}
or, equivalently,
\begin{equation} 
\alpha - \sum _j\frac {t^*(j)-t^*(j+1)}{j-1} 
\le  -\gamma (t) \le \alpha - \sum _j \frac {t^*(j)-t^*(j+1)}{j}   \ . 
\end{equation}
where the sums are limited to the values of $j$ that are reached during the
search. In the large $k$ limit, provided the hypothesis
\begin{equation} \label{hypo}
t^*(j)-t^*(j+1) = \frac 1k + o (1/k)  
\end{equation}
holds for most $j$, we obtain
\begin{equation} \label{re}
-\gamma (t) \simeq \alpha -  \frac 1k \sum _{j^*(t)}^k \frac 1{j}    \ . 
\end{equation}
The hypothesis (\ref{hypo}) is well supported by numerical data, as shown in
figure~\ref{fig_tstar}. 
As the sum of the inverse of the first $k$ integers is equivalent to $\log k$
(harmonic number) we see that the minimal value of $j^*$ over $t$ is
much larger than 2 if $\alpha$ is much smaller than $\log k/k$. Therefore
\begin{equation} \label{alpha_GUC}
\alpha_a \ge \frac {\log k}k \ .
\end{equation}
The r.h.s. of the above inequality coincides with the asymptotic scaling of the
clustering critical ratio (section~\ref{sec:32}). Since the results of the
previous section require that $\a_a \leq \a_d$, we obtain that $\a_a^{\mathrm
(GUC)} =\a_d \simeq \log k/k$ at the leading order in $k\to\io$. As a
comparison, it is easy to see that for UC the threshold for large $k$ is
$\a_a^{\mathrm (UC)} \simeq e/k$, which is therefore much lower than the
threshold for GUC.

These arguments are supported by
numerical simulations that we performed up to $k=2^{16}$, in which the equations
of motion (\ref{eqmot}) are integrated as finite differences equations for all
values of $j$ (see figure~\ref{fig_tstar}).
The numerical investigation confirms that $k \a_a^{\mathrm (GUC)}$ is very well
fitted by $\log k + 2.15$ for $k$ in the range $2^8\div2^{16}$. Moreover, a
finite size scaling analysis (with respect to $k$) of the data shown in
figure~\ref{fig_tstar} shows that 
\begin{equation}
k [t^*(j) - t^*(j+1)] = 1 + k^\nu \times f(j/k)
\end{equation}
where $f(x)$ is a function independent on $k$ which behaves as $x^{-\mu}$ for
$x$ close to 0. From the numerical data, it appears that $\nu = \mu = 0.5$,
which confirms that the first correction to the leading term $\log k/k$ is of
order $1/k$.


\section{Conclusions}
\label{sec:conc}

One of the main results of this paper, that is, that linear-time search
heuristic are not able to solve instances in the clustered phase
of UE-CSP problems
should be interpreted with care. In XORSAT-like models the clustering
transition coincide with the emergence of strong correlations between
variables in solutions, while the two phenomena generally define two
distinct critical ratios for other random decision problems 
\cite{Se07,KZ07}. From an intuitive point of view it is expected that
the performances of search heuristics are affected by correlations
between variables rather than the clustering of solutions. Indeed, as
the search algorithms investigated here do not allow for backtracking
or corrections of wrongly assigned variables, very strong correlations
between $O(N)$ variables (recall that the backbone includes $O(N)$
variables in the clustered phase) are likely to result in $e^{-O(N)}$
probabilities of success for the algorithm.  


Extending the present work to the random Satisfiability ($k$-SAT)
problem would be interesting from this point
of view, because even if the clustering and freezing transition
coincide at leading order for $k\to\io$~\cite{Krzakala07}, their finite $k$
values are different in this case.
Moreover, in some similar problems ($k$-COL~\cite{Achlioptas03} and 1-in-$k$-SAT~\cite{Raymond07})
it has been proven that search algorithms similar to the ones investigated here
are efficient beyond the point where the replica-symmetry-breaking solution
is stable. Therefore these algorithms might beat the clustering threshold in these
problems. Note however that in these cases the transition is continuous, so that
the structure of the clusters is expected to be very different from the one
of XORSAT.

In addition, while the Generalized Unit Clause heuristic is
here shown to be optimal for the $k$-XORSAT problem and to saturate
the clustering ratio when $k\to\infty$, it is certainly not the case
of the $k$-SAT problem. Determining a provably
optimal search heuristic for this
problem remains an open problem.

\vskip1cm


\end{document}